\documentclass[11,twocolumn]{article}
\usepackage{lipsum}
\usepackage{amsmath}
\usepackage{graphicx} 
\usepackage[margin=0.5in]{geometry}
\usepackage{authblk} 

\usepackage[style=ieee]{biblatex} 
\title{Optomagnetism with plasmonic skyrmion}
\author[1]{Vage Karakhanyan}
\author[1,*]{Thierry Grosjean}
\affil[1]{Optics Department, FEMTO-ST Institute UMR 6174, University of Franche-Comté – CNRS\\Besançon, France}
\affil[*]{Corresponding author email: thierry.grosjean@univ-fcomte.fr}

\date{}

\begin{document}
\maketitle
------------------------------  \textbf{Abstract}  ------------------------------\\
Research at the frontier between optics and magnetism is revealing a wealth of innovative phenomena and avenues of exploration.  Optical waves are demonstrating the capacity to induce ultrafast magnetism, while optical analogs of magnetic states, such as magnetic skyrmions, offer the prospect of novel spin-optical states. In this paper, we strengthen the synergy between light and magnetism by exploring the ability of plasmonic Neel skyrmions to create a stationary magnetic field within a thin gold film. We show that, when generated using a focused radially-polarized vortex beam, a plasmonic Neel skyrmion emerges as an optimum for inducing optomagnetism in a thin gold film. Optical skyrmions offer new degrees of freedom for enhancing and controlling optomagnetism in plasmonic nanostructures, with direct application in all-optical magnetization switching, magnetic recording, and the excitation of spin waves.

---------------------------------------------------------------------------\\

Magnetic skyrmions are topologically-protected magnetic states showing space-variant spin distributions \cite{fert_natnano13}. Since their first 
observation in 2009 \cite{muhlbauer2009skyrmion,neubauer2009topological,pappas2009chiral} these spin textures have attracted much interest from both fundamental aspects and potential applications in novel spintronic devices. 

Recently, the concept of magnetic skyrmion has been extended to  optics, leading to novel optical-field or optical-spin textures on the subwavelength scale \cite{tsesses:sci18,du:np19}. As their magnetic counterpart, optical skyrmions of different structures can be densely packed in square or hexagonal lattices \cite{tsesses:sci18,lei:prl21}. Optical skyrmions are a promising building block of the emerging spin-based optics, including optical nano-imaging, quantum information processing, metrology and data storage. 

In this paper, we study the ability of a Neel-type optical skyrmion to generate a stationary magnetic field via the inverse Faraday effect (IFE) \cite{pershan:pr66,popova:prb11,hertel:jmmm06}. The IFE has been extensively explored due to its potential to generate ultrafast magnetic data storage \cite{beaurepaire:prl96,stanciu:prl07,kirilyuk:rmp10} and non-contact excitation of spin-waves \cite{kimel:nature05,kalashnikova:prb08,satoh:natphot12,savochkin:scirep17,matsumoto:prb20}. Moreover, the IFE can be enhanced and tailored via the excitation of surface plasmons in noble metals 
\cite{smolyaninov:prb05,gu:josab10,cheng:np20,koshelev:prb15,hamidi:oc15,nadarajah:ox17,hurst:prb18,mondal:prb15,karakhanyan:ol21,karakhanyan:osac21} 
and in hybrid structures combining noble metals and magnetic materials 
\cite{chu:ome20,ignatyeva:natcomm19,im:prb19,cheng:nl20}. 
Since the IFE is induced by both the SAM and OAM of light, spin-orbit interaction plays a significant role in the optomagnetism \cite{karakhanyan:prb22}. The spin texture of an optical Neel skyrmion, which is also strongly influenced by spin-orbit interaction \cite{du:np19,tsesses:sci18,zayats:npas19}, presents new avenues for generating and controlling optomagnetism. We show here that plasmonic Neel Skyrmion holds promise to maximize optomagnetic effects in a thin gold film. The optical counterpart of magnetic skyrmions offers intriguing solutions to increase and control optomagnetic effects is nanostructures.

\begin{figure}[hbt!]
\centering
\includegraphics[width=1\linewidth]{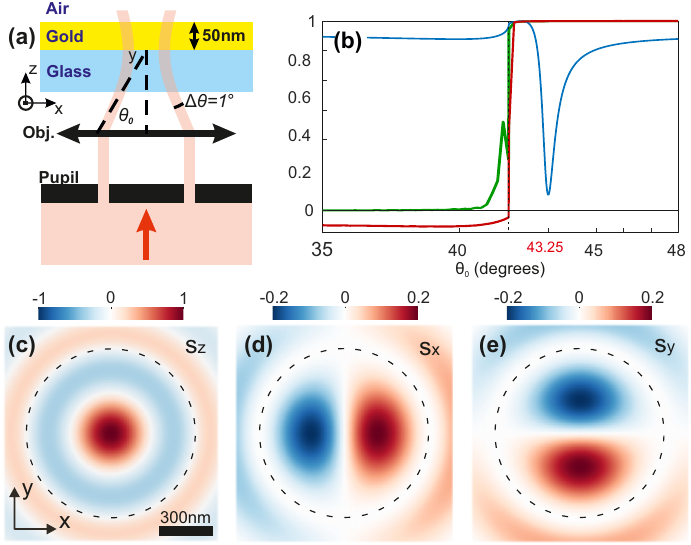}
\caption {(a) Schematic diagram of the focusing system designed to generate a Neel-type plasmonic skyrmion. (b) Blue line: reflectance of the thin gold film as a function of the incidence angle of a $TM$-polarized plane wave in the glass substrate ($\lambda$=800 nm). The vertical dashed line identifies the critical angle of total internal reflection. 
Red and green lines: skyrmion number of the field transmitted through the thin gold film versus the mean incidence angle of the focused field, for the RPVB and the CPB, respectively. The skyrmion number is calculated at a distance of 10 nm beyond the top surface. Focusing occurs across an angular range $\Delta\theta = 0.1^{\circ}$. (c-e) SAM structure of the transmitted optical fields at plasmon resonance, for an incoming RPVB: vector components of the polarization helicity $\boldsymbol{\sigma}$ (Eq. \ref{eq:s2}) at an incidence angle of 43.25$^{\circ}$, with $\Delta \theta = 1^{\circ}$.}
\label{fig:s}
\end{figure}

A plasmonic Neel skyrmion is a metal/air surface mode whose electric field or spin angular momentum (SAM) density is both topologically invariant and radially distributed along the surface. Following approaches introduced by Du et al. \cite{du:np19} and Tsesses et al. \cite{tsesses:sci18}, we generate optical Neel skyrmions by tightly focusing either a radially-polarized vortex beam (RPVB) of topological charge $l=\pm$1  or a circularly polarized beam (CPB) onto a thin gold film lying on a glass substrate (Fig. \ref{fig:s}(a)). RPVB belong to the family of the vector vortex beams, characterized by an inhomogeneous vector polarization state and a helical phase \cite{hao2010phase,zhao2013metamaterials,qiu2014engineering}. 
Focused circularly polarized fields are employed as a standard for inducing IFE 
\cite{pershan:pr66,popova:prb11,hertel:jmmm06,beaurepaire:prl96,stanciu:prl07,kirilyuk:rmp10,smolyaninov:prb05,gu:josab10,cheng:np20,koshelev:prb15,hamidi:oc15,nadarajah:ox17,hurst:prb18,mondal:prb15,karakhanyan:ol21,karakhanyan:osac21}.

The waist of these incoming paraxial beams, characterized by a Gaussian profile, is projected onto the exit pupil plane of a microscope objective where it is spatially filtered with an narrow annular slit within an opaque screen. The annular pupil restricts the incidence angles of incoming light waves to a narrow range around a mean value defined by the slit diameter. This angular range is set either at 1$^{\circ}$ or 0.1$^{\circ}$, regardless of the average incidence angle. The $1/e$ width of the beam waist coincides with the pupil diameter of the microscope objective, whose numerical aperture (NA) of 1.3 enables light focusing in the substrate beyond the critical angle. 


Using the theory established by Richards and Wolf \cite{richards59,novotny:book}, the electric optical field at focus reads:

\begin{equation}
\begin{split}
&\textbf{E}(r,\xi,z)=-\frac{i k_1 f \exp[-i k_1 f]}{2\pi} \frac{1}{\sqrt{n_1}} \times \\ 
&\int^{\theta_0+\frac{\Delta\theta}{2}}_{\theta_0-\frac{\Delta\theta}{2}}
 G(\theta) \int^{2\pi}_0 \textbf{e}(\theta,\psi,z) \exp \left[ i \alpha r \cos (\psi-\xi) \right] d\theta d\psi, \label{eq:sop}
\end{split}
\end{equation}

where $G(\theta) = \cos^{\frac{1}{2}}(\theta)\sin(\theta) F(\theta)$ $(r,\xi,z)$ are cylindrical coordinates,  $f$ is the focal length of the microscope objective, $\theta$ and $\psi$ are directional angles and $\alpha$ is a function of $\theta$. $\Delta\theta$ is the focusing angular range. 

For the RPVB of the first order, the vector $\textbf{e}(\theta,\psi,z)$ takes the form: 

\begin{equation}
\begin{aligned}
\mathbf{e}(\theta,\psi,z) & =\mathbf{e_{TM}}(\theta,\psi,z)\\
  & \propto \exp [\pm i \psi]
\begin{pmatrix}
- t^r_{TM}(z) \cos \theta \cos \psi \\
- t_{TM}^r(z) \cos \theta \sin \psi \\
t_{TM}^z(z) \sin \theta
\end{pmatrix}
\label{eq:field1}
\end{aligned}
\end{equation}

$t_{TM}^r$ and $t_{TM}^z$ are coefficients under the form $C_{TM}^+ \exp[iw_2 z] + C_{TM}^- \exp[-iw_2 z]$ where $w_2$ is the component of the wave vector normal to the surfaces and $i=\sqrt{-1}$. Coefficients $C_{TM}^+$ and $C_{TM}^-$ are obtained by applying boundary conditions of the optical fields at the metal surfaces (see for instance in Ref. \cite{born:book}). 

For a CPB, the vector $\textbf{e}(\theta,\psi,z)$ is written as: 

\begin{equation}
\mathbf{e}(\theta,\psi,z) \propto -\frac{\sqrt {2}}{2}\left[ \mathbf{e_{TM}}(\theta,\psi,z) + i \mathbf{e_{TE}}(\theta,\psi,z) \right],\label{eq:field2}
\end{equation}

where,

\begin{equation}
\mathbf{e_{TE}}(\theta,\psi,z) = t_{TE}(z) \exp [\pm i \psi] 
  \begin{pmatrix}
- \sin \psi \\
 \cos \psi \\
0
\end{pmatrix}.\label{eq:field2b}
\end{equation}

$t_{TE}$  is a coefficient under the form $C_{TE}^+ \exp[iw_2 z] + C_{TE}^- \exp[-iw_2 z]$. Coefficients $C_{TE}^+$ and $C_{TE}^-$ are obtained by applying boundary conditions of the optical fields at the metal surfaces. The phase term $exp [\pm i \psi]$ in Eq. \ref{eq:field2b} reveals spin-orbit interaction at focus \cite{bliokh:sci15}. 

The apodization function at the pupil plane of the microscope objective is approximated by the function:

\begin{equation}
F(\theta)=\frac{2}{w_0}\sqrt{\frac{Z_0 P_0}{\pi}} \exp \left[\frac{- f^2 \sin^2\theta}{w_0^2} \right], \label{eq:apod_gauss}
\end{equation}

with $w_0 = f\sin \theta_M$. \\

$P_0$ and $w_0$ are the power and $1/e$ width of the incoming paraxial beam  and $Z_0$ is the vacuum impedance. The optical magnetic field is calculated by replacing $\textbf{e}(r,\xi,z)$ by $\textbf{h}(r,\xi,z)=\textbf{k}\times\textbf{e}(r,\xi,z) / \omega \mu_0$ in Eq. \ref{eq:sop}. $\omega$ and $\mu_0$ are the angular frequency and permeability of vacuum, respectively. 

At focus, the SAM density of the fields transmitted through the gold film is defined as:

\begin{equation}
\mathbf{s}=\left[ \operatorname{Im}(\varepsilon_0 \mathbf{E} \times \mathbf{E^{*}})+\operatorname{Im}(\mu_0 \mathbf{H} \times \mathbf{H^{*}}) \right] /2\omega, \label{eq:s}
\end{equation}

where $\varepsilon_0$ is the permittivity of vacuum \cite{berry:joa09,bliokh:nc14}. The SAM density can also be expressed as: 

\begin{equation}
\mathbf{s}=\frac{w}{\omega} \boldsymbol{\sigma}, \label{eq:s2}
\end{equation} 

where $w=1/2 \left[ \varepsilon_0 |\mathbf{E}|^2+\mu_0 |\mathbf{H}|^2 \right]$ is the energy density and $\boldsymbol{\sigma}$ is the local polarization helicity vector. The three components of $\boldsymbol{\sigma}$ are comprised between -1 and 1, the values 0 and $\pm 1$ correspond to the linear and circular polarization states, respectively. 

To confirm the analogy to magnetic skyrmions, we calculate the skyrmion number ($n$) associated with the SAM distribution:

\begin{equation}
    n=\dfrac{1}{4\pi}\int_A \mathbf{S_n} \cdot\left(\dfrac{\partial \mathbf{S_n}}{\partial x} \times \dfrac{\partial \mathbf{S_n}}{\partial y} \right) dxdy,
    \label{eq:ch2_skyrmion number}
\end{equation}

where $S_n$ is the unit vector in the direction of the local SAM,  $A$ is the area of a unit cell of the skyrmion.  When $n=1$, the photonic spin structure is the analogue to the magnetization texture of either a Neel or a Bloch skyrmion in magnetic materials \cite{bera2019theory,yu2018transformation}. An analysis of the vector spin distribution allows to identify the exact nature of the optical skyrmion.


On the basis of a hydrodynamic model of the free electrons in a metal, we have recently proposed a new framework to describe the IFE in plasmonic nanostructures \cite{karakhanyan:ol21,karakhanyan:osac21,karakhanyan:prb22}. In the present study, the opto-induced drift current densities within the metal film  are represented as follows:

\begin{equation}\label{eq:jdbulk}
 \begin{split}
 J^b_\xi \approx \frac{2 \mu_0 |\gamma_{\omega}|^2}{ n_0 e} \langle \mathbf{\Pi} \rangle_\xi,
\end{split}
\end{equation}

in the metal bulk, and:

\begin{equation}\label{eq:jdtheta}
J^s_\xi \approx \pm \frac{|\gamma_\omega|^2}{n_0 e \omega}\operatorname{Im} \left[E_{\omega}^z(0^-) E_{\omega}^{\xi \ast}(0^-) \right],
\end{equation}

at the two metal surfaces \cite{karakhanyan:ol21,karakhanyan:prb22}. $\langle \mathbf{\Pi} \rangle_\xi$ is the azimuthal component of the time averaged Poynting vector $\langle \mathbf{\Pi} \rangle = 0.5 \operatorname{Re}\left( \mathbf{E} \times \mathbf{H^*}\right)$. $E_{\omega}^z(0^-)$ and $E_{\omega}^{\xi \ast}(0^-)$ are the longitudinal and azimuthal components of the optical electric field at the two metal surfaces. The $+$ and $-$ sign in Eq. \ref{eq:jdtheta} corresponds to the current densities at the lower and upper interfaces of the metal film, respectively. The optomagnetic field is deduced from Eqs. \ref{eq:jdbulk} and \ref{eq:jdtheta} by using the Biot and Savart law. 




The calculations were conducted at a wavelength of $\lambda = 800$ nm. We identified a surface plasmon excitation at an incidence angle $\theta$ of 43.25$^\circ$ (following the Kretschmann configuration; see Fig. \ref{fig:s}(b)). Figure \ref{fig:s}(b-e) shows an analysis of the vector spin structure generated at a distance of 10 nm beyond the gold film's upper interface by focusing RPVB and CPB. With a skyrmion number of 1 beyond the critical angle of 41.8$^{\circ}$ (Fig. \ref{fig:s}(b)), the photonic spin structures generated at plasmon resonance by both beams are an optical analogue to a magnetic skyrmion. The radially-distributed spin texture observed in Figs. \ref{fig:s}(c-e) for the RPVB identifies a plasmonic Neel skyrmion. A similar spin texture is obtained for circular polarization. Outside the plasmon resonance, below the critical angle, the SAM texture transitions to an azimuthally polarized state, akin to the magnetization texture observed in Bloch skyrmions. The skyrmion number is however insufficient to categorize the optical field as an optical analogue to the Bloch skyrmion.


\begin{figure}[hbt!]
\centering
\includegraphics[width=1\linewidth]{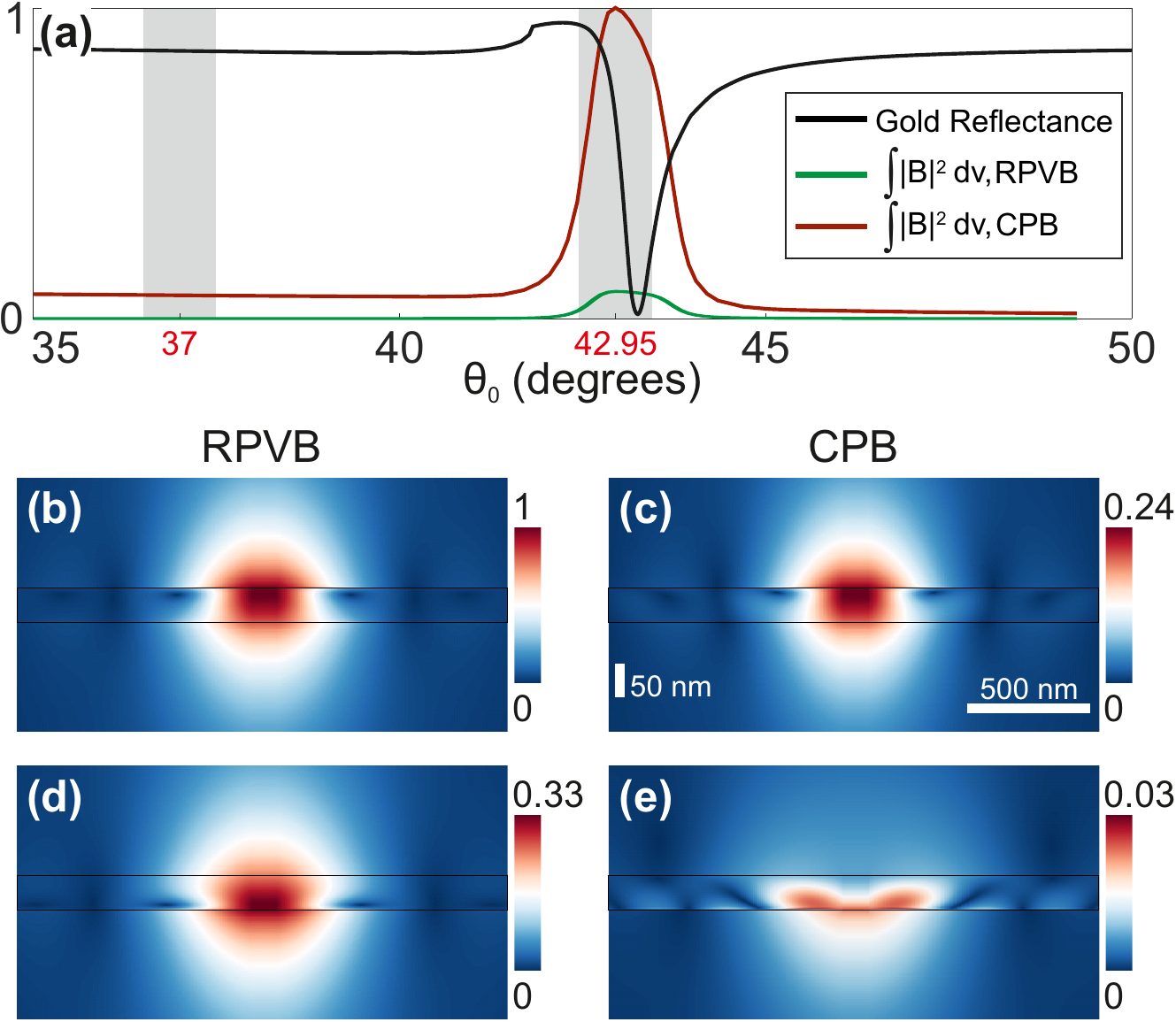}
\caption{(a) Blue curve: Reflectance of the metal film versus the incidence angle of a $TM$-polarized plane wave in the glass substrate ($\lambda$=800 nm). Red and green curves: total energy of the optomagnetic field upon illumination with a focused RPVB (red line) or a CPB  (green line), respectively. (b-e): Amplitude of the optomagnetic field in a longitudinal cross-section, for (b,d) a focused RPVB of the first order, and (c,e) a focused CPB. 
The stationary magnetic fields are calculated (b,c) on resonance and (d,e) off-resonance, with $\Delta\theta=1^{\circ}$ falling within the shaded regions depicted in (a).} \label{fig:B}
 \end{figure}

Fig. \ref{fig:B} shows the optomagnetic response of the thin gold film upon illumination with focused RPVB and CPB, with $\Delta\theta = 1^{\circ}$. Upon scanning the average incidence angle, the total optomagnetic energy (defined in the inset of Fig. \ref{fig:B}(a)) peaks right at the plasmon angle for both incident polarization states, coinciding with the minimum optical reflectance (Fig. \ref{fig:B}(a)). At surface plasmon excitation, the optomagnetic energy is enhanced by approximately 13-fold and 128-fold for RPVB and CPB, respectively. This  confirms the role of surface plasmons in amplifying optomagnetism in metals. However, despite their common skyrmion nature at plasmon resonance, the transmitted fields obtained with the RPVB and CPB result in distinct levels of optomagnetism: the optomagnetic energy is 11.5 times larger with the plasmonic Neel-type skyrmion produced by the RPVB.


Fig. \ref{fig:B}(b) and (c)  show the optomagnetic field generated when a plasmonic Neel skyrmion  is applied to the thin gold film. Using the focused RPVB (Fig. \ref{fig:B}(b)), the maximum optomagnetic field is localized at the upper surface of the gold film, at the surface plasmon resonance.  Off-resonance (Fig. \ref{fig:B}(d)), the optomagnetic field is confined at the lower interface with its maximum reduced by a factor of 3. As a comparison, the optomagnetic fields generated on and off plasmon resonance with the focused CPB are attenuated by 4 and 30 times, respectively (see Figs. \ref{fig:B}(c) and \ref{fig:B}(e), respectively). The optomagnetic fields obtained at plasmon resonance for both polarizations show similar spatial distributions (Figs. \ref{fig:B}(b) and (c)). A similar morphology of optomagnetic field remains off-resonance for the RPVB (cf. Figs. \ref{fig:B}(b) and (d)), but is lost with the CPB (cf. Figs. \ref{fig:B}(c) and (e)). 

\begin{figure}[hbt!]
\centering
\includegraphics[width=1\linewidth]{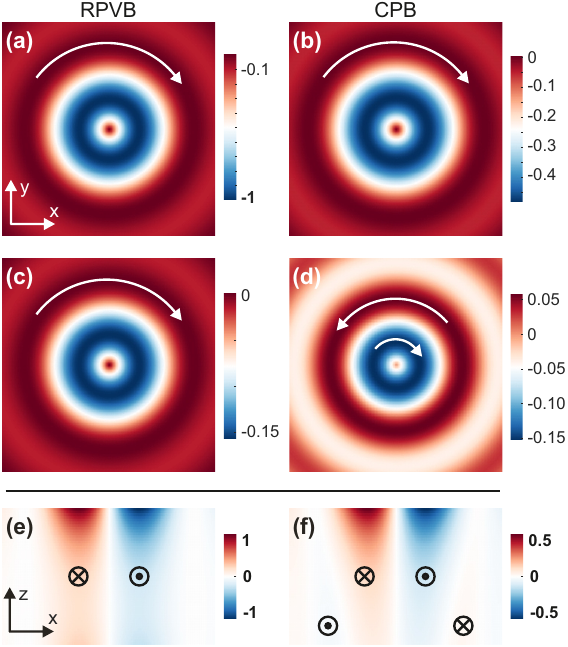}
\caption {(a-d) Normalized current densities at (a,b) the upper and (c,d) lower surfaces of the thin gold film, with (a,c) a focused RPVB and (b,d) a focused CPB. (e,f) Normalized volume current densities in a (xz) cross-section perpendicular to the surfaces, with focused RPVB and CPB, respectively.}
\label{fig:current_res}
\end{figure}


According to Eqs. \ref{eq:sop} and \ref{eq:field1}, a focused RPVB generates an axis-symmetrical "$TM$ optical state" within the metal film,  which is constituted of $TM$-polarized Fourier components. In contrast, when a focused CPB is applied, a combination of axis-symmetrical $TE$ and $TM$ optical states arises within the metal film (see Eqs. \ref{eq:sop}, \ref{eq:field2} and \ref{eq:field2b}). The "$TE$ optical state", constituted of $TE$-polarized Fourier components, is primarily confined within the skin depth at the lower interface. 

In the case of the RPVB, the entirety of the incident field intensity contributes to the generation of the $TM$ optical state within the metal film, which includes surface plasmons (cf. Eq \ref{eq:field1}). By comparing Eqs. \ref{eq:field1} and \ref{eq:field2}, we see that only half of the incident field intensity is involved for circular polarization. As the surface and volume current densities stem from a second-order nonlinear optical effect (see Eqs. \ref{eq:jdbulk} and \ref{eq:jdtheta}) \cite{karakhanyan:ol21}, the plasmonic Neel skyrmion produced with the RPVB exhibits twice the efficiency in generating opto-induced drift currents compared to the one obtained with the CPB, as shown in Fig. \ref{fig:current_res}. Since  the $TE$ optical state is confined within the lower skin depth, the plasmon-induced drift current densities exhibit similar spatial distributions in the upper half of the gold film, as shown in Figs. \ref{fig:current_res}(a,b) and \ref{fig:current_res}(e,f). However, circular polarization results in a maximum amplitude of the drift current densities that is halved.

With the focused RPVB, the plasmonic Neel skyrmion generates surface and volume drift current densities of the same handedness, forming a uniform current loop (see Figs. \ref{fig:current_res}(a,c,e)). In contrast, the presence of a $TE$ optical state with the CPB induces a pair of current loops with opposite handedness in the lower half of the metal film, which is less efficient in generating an optomagnetic field (Figs. \ref{fig:current_res}(d,f)). Therefore, the existence of  $TE$ optical states in the plasmonic film both attenuates amplitude of the plasmon field and hinders the optomagnetic effect in the lower half of the metal film. As a result, in our study, the optomagnetic field is reduced by four with the focused CPB compared to the focused RPVB (which lacks $TE$ optical state), although an optical Neel skyrmion is transmitted through the metal film in both cases. Therefore, by creating a pure $TM$ optical state in a thin gold film, the plasmonic Neel skyrmion obtained with the RPVB emerges as an optimum for generating plasmon-induced magnetism. 

To conclude, an optical counterpart of magnetic Neel skyrmion is shown to generate a static magnetic field in a thin gold film, via plasmon-enhanced IFE. We evidence that circular polarization, the standard for producing IFE, is not the optimal choice for inducing plasmon-enhanced optomagnetism in a thin gold film. The optimum is achieved with a plasmonic Neel skyrmion obtained from a vector vortex beam. Vector vortex beams, which produce optical skyrmions by exploiting spin-orbit interaction, open new avenues for generating and controlling plasmon-induced magnetism in metal films, and by extension, in resonant plasmonic nanoantennas. 

\section*{Funding}

EIPHI Graduate School (contract ANR-17-EURE-0002); Region "Bourgogne Franche-Comte"; French Agency of Research (contracts ANR-18-CE42-0016 and ANR-23-CE42-0021); French Renatech network; Equipex+ Nanofutur (21-ESRE-0012)

\section*{Disclosures}
\medskip
\noindent The authors declare no conflicts of interest.

 \printbibliography
\end{document}